\RequirePackage{ifpdf}
\ifpdf 
\documentclass[pdftex]{sigma}
\else
\documentclass{sigma}
\fi

\newcommand{\half}{\mbox{$\textstyle\frac{1}{2}$}}
\newcommand{\cP}{{\cal P}}
\newcommand{\cT}{{\cal T}}
\newcommand{\cC}{{\cal C}}
\newcommand{\cQ}{{\cal Q}}

\begin{document}

\renewcommand{\PaperNumber}{126}

\FirstPageHeading

\renewcommand{\thefootnote}{$\star$}

\ShortArticleName{Faster than Hermitian Time Evolution}

\ArticleName{Faster than Hermitian Time Evolution\footnote{This
paper is a contribution to the Proceedings of the Seventh
International Conference ``Symmetry in Nonlinear Mathematical
Physics'' (June 24--30, 2007, Kyiv, Ukraine). The full collection
is available at
\href{http://www.emis.de/journals/SIGMA/symmetry2007.html}{http://www.emis.de/journals/SIGMA/symmetry2007.html}}}

\Author{Carl M. BENDER}
\AuthorNameForHeading{C.M. Bender}

\Address{Physics Department, Washington University, St. Louis, MO 63130, USA}
\Email{\href{mailto:cmb@wustl.edu}{cmb@wustl.edu}}
\URLaddress{\url{http://www.physics.wustl.edu/~cmb/}}

\ArticleDates{Received October 22, 2007, in f\/inal form December 22, 2007; Published online December 26, 2007}

\Abstract{For any pair of quantum states, an initial state $|I\rangle$ and a
f\/inal quantum state~$|F\rangle$, in a Hilbert space, there are many Hamiltonians
$H$ under which $|I\rangle$ evolves into $|F\rangle$. Let us impose the
constraint that the dif\/ference between the largest and smallest eigenvalues of~$H$, $E_{\max}$ and $E_{\min}$, is held f\/ixed. We can then determine the
Hamiltonian $H$ that satisf\/ies this constraint and achieves the transformation
from the initial state to the f\/inal state in the least possible time $\tau$. For
Hermitian Hamiltonians, $\tau$ has a nonzero lower bound. However, among
non-Hermitian $\cP\cT$-symmetric Hamiltonians satisfying the same energy
constraint, $\tau$~can be made arbitrarily small without violating the
time-energy uncertainty principle. The minimum value of $\tau$~can be made
arbitrarily small because for $\cP\cT$-symmetric Hamiltonians the path from the
vector $|I\rangle$ to the vector $|F\rangle$, as measured using the
Hilbert-space metric appropriate for this theory, can be made arbitrarily short.
The mechanism described here is similar to that in general relativity in which
the distance between two space-time points can be made small if they are
connected by a wormhole. This result may have applications in quantum
computing.}


\Classification{81Q10; 81S99}

\section{Classical brachistochrone problem}

Three hundred years ago the solution to a famous problem in classical mechanics
known as the {\it brachistochrone} was found almost simultaneously by a number
of distinguished mathematicians including Newton, Bernoulli, Leibniz, and
L'H\^opital. (The term {\it brachistochrone} is derived from Greek and means
{\it shortest time}.) The problem is stated as follows: A bead slides down a~frictionless wire from one given point to another in a homogeneous gravitational
f\/ield. What is the shape of the wire connecting the two points that minimizes
the time of descent of the bead? The solution found by these mathematicians is
that the wire must be in the shape of a~cycloid.

Of course, it is implicitly assumed in the derivation of the brachistochrone
that the path of shortest time of descent is {\it real}. It is interesting that
if one allows for the possibility of complex paths of motion, one can achieve
an even shorter time of f\/light.

\begin{figure*}[!th]\centering
\includegraphics{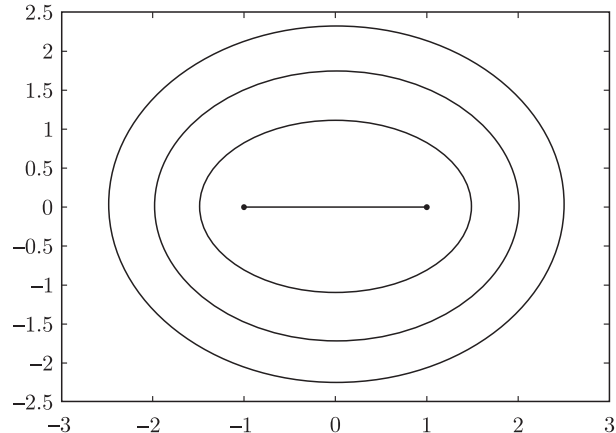}
\caption{Classical trajectories in the complex-$x$ plane for the harmonic
oscillator whose Hamiltonian is $H=p^2+x^2$. These trajectories represent the
possible paths of a particle whose energy is $E=1$. The trajectories are nested
ellipses with foci located at the turning points at $x=\pm1$. The real line
segment (degenerate ellipse) connecting the turning points is the usual periodic
classical solution to the harmonic oscillator. All closed paths have the same
period $2\pi$.}
\label{f1}
\end{figure*}

To illustrate how shorter times can be achieved by means of complex paths, let
us consider the simple classical harmonic oscillator, whose Hamiltonian is given
by
\begin{gather*}
H=p^2+x^2.
\end{gather*}
If we have a particle of energy $E=1$, then the classical turning points of the
motion of the particle are located at $x=\pm1$. The particle undergoes simple
harmonic motion in which it oscillates sinusoidally between these two turning
points. This periodic motion is indicated in Fig.~\ref{f1} by a solid line
connecting the turning points. However, in addition to this oscillatory motion
on the real-$x$ axis, there are an inf\/inite number of other trajectories that a
particle of energy $E$ can have \cite{Bender1999}. These classical trajectories,
which are also shown in Fig.~\ref{f1}, are all ellipses whose foci are located
at precisely the positions of the turning points. All of the classical orbits
are periodic and all orbits have the same period $T=2\pi$. Thus, a classical
particle travels faster along more and more distant ellipses.

Now suppose that a classical particle of energy $E=1$ is traveling along the
real-$x$ axis from some point $x=-a$ to $x=a$, where $a>1$. If the potential $V$
is everywhere zero along its path, then it will travel at a constant velocity.
However, if the particle suddenly f\/inds itself in the parabolic potential $V(x)=
x^2$ just as it reaches the turning point at $x=-1$ and it suddenly escapes the
inf\/luence of this potential at $x=1$, then the time of f\/light from $x=-a$ to
$x=a$ will be changed because the particle is not traveling at constant velocity
between the turning points. Now imagine that the potential $V(x)=x^2$ is
suddenly turned on {\it before} the particle reaches the turning point at $x=-
1$. In this case, the particle will follow one of the elliptical paths in the
complex plane around to the positive real axis. Just as the particle reaches
the positive real axis the potential is turned of\/f, so the particle proceeds
onward along the real axis until it reaches $x=a$. This trip will take less
time because the particle travels faster along the ellipse in the complex plane.

We have arrived at the surprising conclusion that if the classical particle
enters the parabolic potential $V(x)=x^2$ immediately after it begins its voyage
up the real axis, its time of f\/light will be exactly half a period, or $\pi$.
Indeed, by traveling in the complex plane, a particle of energy $E=1$ can go
from the point $x=-a$ to the point $x=a$ in time $\pi$, no matter how large $a$
is. Evidently, if a particle is allowed to follow complex classical
trajectories, then it is possible to make a drastic reduction in its time of
f\/light between two given real points.

\section{Quantum brachistochrone problem}

The purpose of this paper is to show that by using complex non-Hermitian
Hamiltonians, we can obtain a faster time of f\/light than is possible with
Hermitian Hamiltonians. The quantum brachistochrone problem is def\/ined as
follows: If we are given an initial quantum state $|I\rangle$ and a~f\/inal
quantum state $|F\rangle$, then there exist many Hamiltonians $H$ under which
$|I\rangle$ evolves into~$|F\rangle$. The quantum brachistochrone problem is to
f\/ind the particular Hamiltonian $H$ that achieves this transformation in the
least time $\tau$, subject to the constraint that the dif\/ference between the
largest and smallest eigenvalues of $H$ is held f\/ixed. For Hermitian
Hamiltonians, $\tau$ has a nonzero lower bound. However, we will see that for
non-Hermitian $\cP\cT$-symmetric Hamiltonians satisfying the same energy
constraint, $\tau$ can be made arbitrarily small.

One might think that this result could violate the time-energy uncertainty
principle. However, we will see that this is not the case because for
non-Hermitian Hamiltonians the path from $|I\rangle$ to $|F\rangle$ can be made
arbitrarily short. The mechanism is similar to that in general relativity, where
the distance between two space-time points can be made small if they are
connected by a wormhole.

\section[Review of $PT$ quantum mechanics]{Review of $\boldsymbol{\cP\cT}$ quantum mechanics}

Based on the traditional training that one receives in a quantum mechanics
course, one would expect a theory def\/ined by a non-Hermitian Hamiltonian to be
unphysical because the energy levels would most likely be complex and the time
evolution would most likely be nonunitary (not probability-conserving). However,
theories def\/ined by a special class of non-Hermitian Hamiltonians called $\cP
\cT$-symmetric Hamiltonians can have positive real energy levels and can exhibit
unitary time evolution. Such theories are acceptable quantum theories. In
principle, these theories can be distinguished experimentally from those def\/ined
by Hermitian Hamiltonians because non-Hermitian time evolution can proceed
arbitrarily rapidly.

We use the following notation in this paper: By the term {\it Hermitian}, we
mean {\it Dirac} Hermitian, where the Dirac Hermitian adjoint symbol $^\dag$
represents combined matrix transposition and complex conjugation. The parity
operator $\cP$ performs spatial ref\/lection $x\to-x$ and the antilinear
time-reversal operator $\cT$ performs combined time reversal $t\to-t$ and
complex conjugation.

The f\/irst $\cP\cT$-symmetric quantum-mechanical Hamiltonians were introduced in
1998 \cite{Bender1998}; and since then there have been many follow-up papers by
a wide range of authors. There have also been three recent review articles
\cite{Bender2005,Bender2007,Dorey2007}. In~\cite{Bender1998} it was
discovered that even if a Hamiltonian is not Hermitian, its energy levels can be
all real and positive so long as the eigenfunctions are symmetric under $\cP\cT$
ref\/lection.

These new kinds of Hamiltonians are obtained by deforming ordinary Hermitian
Hamiltonians into the complex domain. The original class of $\cP\cT$-symmetric
Hamiltonians that was proposed in~\cite{Bender1998} has the form
\begin{gather}
\label{e1}
H=p^2+x^2(ix)^\epsilon\qquad(\epsilon>0),
\end{gather}
where $\epsilon$ is a real deformation parameter. Two particularly interesting
special cases are obtained by setting $\epsilon=1$ to obtain $H=p^2+ix^3$ and
by setting $\epsilon=2$ to obtain $H=p^2-x^4$. Surprisingly, these Hamiltonians
have real, positive, discrete energy levels even though the potential for
$\epsilon=1$ is imaginary and the potential for $\epsilon=2$ is upside-down. The
f\/irst proof of spectral reality and positivity for $H$ in (\ref{e1}) was given
by Dorey {\it et al.} in~\cite{Dorey2001}.

The philosophical background of $\cP\cT$ quantum mechanics is simply this: The
axiom of quantum mechanics that requires the Hamiltonian $H$ to be Dirac
Hermitian is distinct from all of the other axioms because it is mathematical in
character rather than physical. The other axioms are stated in physical terms;
these other axioms require locality, causality, stability and uniqueness of the
vacuum state, conservation of probability, Lorentz invariance, and so on. The
condition of Dirac Hermiticity $H=H^\dag$ is mathematical, but the condition of
$\cP\cT$ symmetry $H=H^{\cP\cT}$ (space-time ref\/lection symmetry) is physical
because $\cP$ and $\cT$ are elements of the Lorentz group.

The spectrum $H$ in (\ref{e1}) is real, which poses the question of whether this
Hamiltonian specif\/ies a {\it quantum-mechanical} theory. That is, is the theory
specif\/ied by $H$ associated with a Hilbert space endowed with a positive inner
product and does $H$ specify unitary (norm-preserving) time evolution? The
answer to these questions is {\it yes}. Positivity of the inner product and
unitary time evolution was established in~\cite{Bender2002} for
quantum-mechanical systems having an unbroken $\cP\cT$ symmetry and in
\cite{Bender2004} for quantum f\/ield theory.

To demonstrate that the theory specif\/ied by the $H$ in (\ref{e1}) is a
quantum-mechanical theory, we construct a linear operator $\cC$ that satisf\/ies
the three simultaneous algebraic equations \cite{Bender2002}: $\cC^2=1$, $[\cC,
\cP\cT]=0$, and $[\cC,H]=0$. Using $\cC$, which in quantum f\/ield theory is a
Lorentz scalar~\cite{BenderBrandtChenWang2005}, we can then construct the
appropriate inner product for a $\cP\cT$-symmetric Hamiltonian: $\langle a|b
\rangle\equiv a^{\cC\cP\cT}\cdot b$. This inner product, which uses the $\cC\cP
\cT$ adjoint, has a strictly positive norm: $\langle a|a\rangle>0$. Because~$H$
commutes with both $\cP\cT$ and $\cC$, $H$ is {\it self-adjoint} with respect to
$\cC\cP\cT$ conjugation. Also, the time-evolution operator $e^{-iHt}$ is unitary
with respect to $\cC\cP\cT$ conjugation. Note that the Hilbert space and the
$\cC\cP\cT$ inner product is {\it dynamically determined} by the Hamiltonian
itself.

We have explained why a $\cP\cT$-symmetric Hamiltonian gives rise to a unitary
theory, but in~doing so we raise the question of whether $\cP\cT$-symmetric
Hamiltonians are useful. The answer to this question is simply that $\cP
\cT$-symmetric Hamiltonians have {\it already} been useful in~many areas of
physics. For example, in 1959 Wu showed that the ground state of a Bose system
of hard spheres is described by a non-Hermitian Hamiltonian \cite{Wu1959}. Wu
found that the ground-state energy of this system is real and he conjectured
that all of the energy levels were real. Hollowood showed that the non-Hermitian
Hamiltonian for a complex Toda lattice has real energy levels~\cite{Hollowood1992}. Cubic non-Hermitian Hamiltonians of the form $H=p^2+ix^3$
(and also cubic quantum f\/ield theories having an imaginary self-coupling term)
arise in studies of the Lee--Yang edge singularity~\cite{Fisher1978} and in
various Reggeon f\/ield-theory models~\cite{Brower1978}. In all of these cases a
non-Hermitian Hamiltonian having a real spectrum appeared mysterious at the
time, but now the explanation is simple: In every case the non-Hermitian
Hamiltonian is $\cP\cT$ symmetric. Hamil\-tonians having $\cP\cT$ symmetry have
also been used to describe magnetohydrodynamic sys\-tems~\cite{Guenther2005} and
to study nondissipative time-dependent systems interacting with electromagnetic
f\/ields~\cite{Fring2006}.

An important application of $\cP\cT$ quantum mechanics is in the revitalization
of theories that have been abandoned because they appear to have ghosts. {\it
Ghosts} are states having negative norm. We have explained above that in order
to construct the quantum-mechanical theory def\/ined by a $\cP\cT$-symmetric
Hamiltonian, we must construct the appropriate adjoint from the $\cC$ operator.
Having constructed the $\cC\cP\cT$ adjoint, one may f\/ind that the so-called
ghost state is actually not a ghost at all because when its norm is calculated
using the correct def\/inition of the adjoint, the norm turns out to be
positive. This is precisely what happens in the case of the Lee model.

The Lee model was proposed in 1954 as a quantum f\/ield theory in which mass,
wave-function, and charge renormalization could be performed exactly and in
closed form~\cite{Lee1954}. However, in 1955 K\"all\'en and Pauli showed that
when the renormalized coupling constant is larger than a~critical value, the
Hamiltonian becomes non-Hermitian (in the Dirac sense) and a ghost state appears~\cite{Kallen1955}. The appearance of the ghost was assumed to be a fundamental
defect of the Lee model. However, the non-Hermitian Lee-model Hamiltonian is
$\cP\cT$ symmetric and when the norms of the states of this model are determined
using the $\cC$ operator, which can be calculated in closed form, the ghost
state is seen to be an ordinary physical state having positive norm~\cite{BenderBrandt2005}. Thus, the following words by Barton~\cite{Barton1963}
are {\it not true}: ``A non-Hermitian Hamiltonian is unacceptable partly because
it may lead to complex energy eigenvalues, but chief\/ly because it implies a
non-unitary S matrix, which fails to conserve probability and makes a hash of
the physical interpretation.''

Another example of a quantum model that was thought to have ghost states, but in
fact does not, is the Pais--Uhlenbeck oscillator model \cite{BenderMannheim2007}.
This model has a fourth-order f\/ield equation, and for the past several decades
it was thought (incorrectly) that all such higher-order f\/ield equations lead
inevitably to ghosts. Indeed, it is explained in~\cite{BenderMannheim2007}
when the Pais--Uhlenbeck model is quantized using the methods of $\cP\cT$ quantum
mechanics, it does not have any ghost states at all.

There are many potential applications for $\cP\cT$ quantum mechanics in areas
such as particle physics, cosmology, gravitation, quantum f\/ield theory, and
solid-state physics. These applications are discussed in detail in the recent
review article~\cite{Bender2007}.

Having established the validity and potential usefulness of $\cP\cT$ quantum
mechanics, one may ask why $\cP\cT$ quantum mechanics works. The reason is that
$\cC\cP$ is a positive operator, and thus it can be written as the exponential
of another operator $\cQ$: $\cC\cP=e^\cQ$. The square root of $e^\cQ$ can then
be used to construct a new Hamiltonian $\tilde H$ via a similarity
transformation on the $\cP\cT$-symmetric Hamiltonian $H$: $\tilde H\equiv e^{-
\cQ/2}He^{\cQ/2}$. The new Hamiltonian $\tilde H$ has the same energy
eigenvalues as the original Hamiltonian $H$ because a similarity transformation
is isospectral. Moreover, $\tilde H$ is {\it Dirac Hermitian}
\cite{Mostafazadeh2003}. $\cP\cT$ quantum mechanics works because there is an
equivalence between a non-Hermitian $\cP\cT$-symmetric Hamiltonian and a
conventional Dirac Hermitian Hamiltonian.

There are a number of elementary examples of this equivalence, but a nontrivial
illustration is provided by the Hamiltonian $H$ in (\ref{e1}) at $\epsilon=2$,
which is not Hermitian because boundary conditions that violate the $L^2$ norm
must be imposed in Stokes wedges in the complex plane in order to obtain a real,
positive, discrete spectrum. The exact equivalent Hermitian Hamiltonian is
$\tilde H=p^2+4x^4-2\hbar\,x$, where $\hbar$ is Planck's constant
\cite{Buslaev1993}. The term proportional to $\hbar$ vanishes in the classical
limit and is thus an example of a quantum anomaly.

We have established that $\cP\cT$ symmetry is equivalent by means of a
similarity transformation to conventional Dirac Hermiticity. Therefore, one may
wonder whether $\cP\cT$ quantum mechanics is actually fundamentally dif\/ferent
from ordinary quantum mechanics? The answer is {\it yes}, and this paper argues
that, at least in principle, there is an experimentally observable dif\/ference
between $\cP\cT$-symmetric and ordinary Dirac Hermitian Hamiltonians. The
quantum brachistochrone provides a setting for examining this dif\/ference and
provides a way to discriminate between the class of $\cP\cT$-symmetric
Hamiltonians and the class of Dirac Hermitian Hamilto\-nians.

\section{Solving the Hermitian quantum brachistochrone problem}

To f\/ind the Hermitian Hamiltonian $H$ that solves the quantum brachistochrone
problem we must examine all possible Hamiltonians under which a state $|\psi_I
\rangle$ in Hilbert space evolves into another state $|\psi_F\rangle$ in time
$t$:
\begin{gather*}
|\psi_I\rangle\to|\psi_F\rangle=e^{-iHt/\hbar}|\psi_I\rangle.
\end{gather*}
The problem is to f\/ind the minimum time $t=\tau$ required for this
transformation, subject to the constraint that the dif\/ference $\omega$ between
the largest and smallest eigenvalues of $H$,
\begin{gather*}
\omega=E_{\max}-E_{\min},
\end{gather*}
is held f\/ixed. The quantum brachistochrone is the Hamiltonian that performs this
time evolution in the least possible time. In~\cite{Carlini2006} it is
shown that for Hermitian Hamiltonians $\tau\neq0$. However, we show in this
paper that one can f\/ind a Hamiltonian in the space of $\cP\cT$-symmetric
Hamiltonians that satisf\/ies the same energy constraint and can perform the time
evolution in no time at all!

Here, we study the simplest case of Hamiltonians having only two energy levels.
We restrict the discussion to this case because it is shown in~\cite{Brody2006} that one need only work in the two-dimensional subspace of
the full Hilbert space that is spanned by the initial state vector $|\psi_I
\rangle$ and the f\/inal state vector $|\psi_F\rangle$. We consider the case of
Hermitian Hamiltonians and choose a~basis so that
\begin{gather*}
|\psi_I\rangle=\left(\begin{array}{c}1\\0\end{array}\right)\qquad{\rm and}\qquad
|\psi_F\rangle=\left(\begin{array}{c}a\\ b\end{array}\right),
\end{gather*}
where the condition that $|\psi_F\rangle$ be normalized is $|a|^2+|b|^2=1$. The
most general $2\times2$ Hermitian Hamiltonian is
\begin{gather*}
H=\left(\begin{array}{cc} s & re^{-i\theta}\cr re^{i\theta}&u\end{array}\right)
\qquad(r,~s,~u,~\theta~{\rm real}).
\end{gather*}
For this Hamiltonian the eigenvalue constraint takes the form
\begin{gather}
\label{e4}
\omega^2=(s-u)^2+4r^2.
\end{gather}

To f\/ind the optimal Hamiltonian satisfying this constraint, we express $H$ in
terms of the Pauli matrices:
\begin{gather*}
H=\half(s+u){\bf 1}+\half\omega\sigma\!\cdot\!{\bf n},
\end{gather*}
where
\begin{gather*}
{\bf n}=\frac{1}{\omega}(2r\cos\theta,2r\sin\theta,s-u)
\end{gather*}
is a unit vector and
\begin{gather*}
\sigma_1=\left(\begin{array}{cc}0&1\\ 1&0\end{array}\right),\qquad
\sigma_2=\left(\begin{array}{cc}0&-i\\ i&0\end{array}\right),\qquad
\sigma_3=\left(\begin{array}{cc}1&0\\0&-1\end{array}\right).
\end{gather*}
We use the identity
\begin{gather*}
\exp(i\phi\,\sigma\!\cdot\!{\bf n})=\cos\phi\,{\bf 1}+i\sin\phi\,\sigma\!
\cdot\!{\bf n}
\end{gather*}
to write $|\psi_F\rangle=e^{-iH\tau/\hbar}|\psi_I\rangle$ as
\begin{gather*}
\left(\begin{array}{c}a\\b\end{array}\right)=e^{-\frac{1}{2}i(s+u)t/\hbar}\left(
\begin{array}{c}\cos\frac{\omega t}{2\hbar}-i\frac{s-u}{\omega}\sin\frac{\omega
t}{2\hbar} \vspace{2mm}\\ -i\frac{2r}{\omega}e^{i\theta}\sin\frac{\omega t}{2\hbar}
\end{array}\right).
\end{gather*}
The second component of this equation then gives $|b|=\frac{2r}{\omega}\sin
\frac{\omega t}{2\hbar}$, which allows us to f\/ind the required time of
evolution:
\begin{gather*}
t=\frac{2\hbar}{\omega}\arcsin\frac{\omega|b|}{2r}.
\end{gather*}

We must now minimize the time $t$ over all $r>0$ while maintaining the
constraint in (\ref{e4}). This constraint tells us that the maximum value of $r$
is $\half\omega$. At this maximum we have $s=u$. The minimum evolution time
$\tau$ is thus given by
\begin{gather}
\label{e6}
\tau\omega=2\hbar\arcsin|b|.
\end{gather}
Note that if $a=0$ and $b=1$, we have $\tau=\pi\hbar/\omega$ for the smallest
time required to transform $\left(1\atop0\right)$ to the orthogonal state
$\left(0\atop1\right)$. The time $\tau$ required to transform a vector into an
orthogonal vector is called the {\it passage time}.

The form of the result in (\ref{e6}) resembles the uncertainty principle, but
(\ref{e6}) is merely the statement that {\it rate$\,\times\,$time$\,=\,$distance}. The
constraint in (\ref{e4}) is equivalent to a bound on the standard deviation
$\Delta H$, where $(\Delta H)^2=\langle\psi|H^2|\psi\rangle-\langle\psi|H|\psi
\rangle^2$ in the normalized state $|\psi\rangle$. The maximum of $\Delta H$ is
$\omega/2$. The {\it speed} of evolution of a quantum state is given by $\Delta
H$. The {\it distance} between the initial state $|\psi_I\rangle$ and the f\/inal
state $|\psi_F\rangle$ is $2\arccos(|\langle\psi_F|\psi_I\rangle|)$. Thus, the
time $\tau$ to evolve from $|\psi_I\rangle$ to $|\psi_F\rangle$ is bounded below
because the speed is bounded above with the distance held f\/ixed.

\section{Solving the non-Hermitian quantum brachistochrone problem}

For a $\cP\cT$-symmetric Hamiltonian, $\tau$ can be arbitrarily small. This is
because a $\cP\cT$-symmetric Hamiltonian whose eigenvalues are all real is
equivalent to a Hermitian Hamiltonian via $\tilde H=e^{-\cQ/2}He^{\cQ/2}$. The
states in a $\cP\cT$-symmetric theory are mapped by $e^{-\cQ/2}$ to the
correspon\-ding states in the Dirac Hermitian theory. But, the overlap distance
between two states does not remain constant under a similarity transformation.
We can exploit this property of the similarity transformation to overcome the
Hermitian lower limit on the time $\tau$. The detailed calculation is explained
in~\cite{BenderBrodyJonesMeister2007}, and this calculation has already led
to much research activity and lively debate~\cite{Assis2007}.

We consider the general class of $\cP\cT$-symmetric $2\times2$ Hamiltonians
having the form
\begin{gather}
\label{e7}
H=\left(\begin{array}{cc}re^{i\theta}&s\cr s & re^{-i\theta}\end{array}\right)
\qquad(r,~s,~\theta~{\rm real}),
\end{gather}
where $\cT$ is complex conjugation and $\cP=\left({0~~1}\atop{1~~0}\right)$. The
eigenvalues
\begin{gather*}
E_\pm =r\cos\theta\pm\sqrt{s^2-r^2\sin^2\theta}
\end{gather*}
are real if $s^2>r^2 \sin^2\theta$. This inequality def\/ines the region of
unbroken $\cP\cT$ symmetry. The unnormalized eigenstates of $H$ are
\begin{gather*}
|E_+\rangle=\left(\begin{array}{c}e^{i\alpha/2} \cr e^{-i \alpha/2}\end{array}
\right),\qquad |E_-\rangle = \left(\begin{array}{c} i e^{-i \alpha/2}\cr -i
e^{i \alpha/2}\end{array} \right),
\end{gather*}
where $\alpha$ (real) is given by $\sin\alpha=(r/s)\sin\theta$. The $\cC$
operator for $H$ in (\ref{e7}) is
\begin{gather*}
\cC=\frac{1}{\cos\alpha}\left(\begin{array}{cc} i\sin\alpha & 1 \cr 1 & -i
\sin\alpha \end{array}\right).
\end{gather*}
It is easy to verify that the $\cC\cP\cT$ norms of both eigenstates have the
value $\sqrt{2\cos\alpha}$.

To calculate $\tau$ we express the $H$ in (\ref{e7}) as
\begin{gather*}
H=(r\cos\theta){\bf 1}+\half\omega\sigma\!\cdot\!{\bf n},
\end{gather*}
where
\begin{gather*}
{\bf n}=\frac{2}{\omega}(s,0,ir\sin\theta)
\end{gather*}
is a unit vector. The squared dif\/ference between energy eigenvalues is{\samepage
\begin{gather}
\label{e10}
\omega^2=4s^2-4r^2\sin^2\theta.
\end{gather}
The positivity of $\omega^2$ is ensured by the condition of unbroken $\cP\cT$
symmetry.}

To determine $\tau$ we write down the $\cP\cT$-symmetric time-evolution
equation:
\begin{gather*}
e^{-iHt/\hbar}\left(\begin{array}{c} 1\\0\end{array}\right)=\frac{e^{-itr\cos
\theta/\hbar}}{\cos\alpha}\left(\begin{array}{c}\cos(\frac{\omega t}{2\hbar}-
\alpha) \vspace{1mm}\\ -i\sin\left( \frac{\omega t}{2\hbar}\right)\end{array}\right).
\end{gather*}
Consider the pair of vectors used in the Hermitian case: $|\psi_I\rangle=\left(1
\atop0\right)$ and $|\psi_F\rangle=\left(0\atop1\right)$. (Note that these two
vectors are not orthogonal with respect to the $\cC\cP\cT$ inner product.)
Observe that the evolution time needed to reach $|\psi_F\rangle$ from $|\psi_I
\rangle$ is $t=(2\alpha-\pi)\hbar/\omega$. Optimizing this result over allowable
values for $\alpha$ as $\alpha$ approaches $\half\pi$, the optimal time $\tau$
tends to zero!

\section{Discussion}

Equations (\ref{e4}) and (\ref{e10}) reveal the dif\/ference between Hermitian and
$\cP\cT$-symmetric Hamiltonians. Equation (\ref{e4}) for the Hermitian matrix
Hamiltonian has a {\it sum} of squares while (\ref{e10}) has a~{\it difference}
of squares. The elliptic equation~(\ref{e4}) gives a nonzero lower bound for
$\tau$. The hyperbolic equation~(\ref{e10}) allows $\tau$ to approach zero
because the matrix elements of a $\cP\cT$-symmetric Hamiltonian can be made
large without violating the energy constraint $E_+-E_-=\omega$. The fact that
$\tau$ can be made arbitrarily small may have applications in quantum computing.

We conclude with two comments. First, as $\alpha\to\half\pi$ we get $\cos\alpha
\to0$. However, the energy constraint becomes $\omega^2=4s^2\cos^2\alpha$. Since
$\omega$ is f\/ixed, to have $\alpha$ approach $\half\pi$, we must require $s\gg
1$. It follows from $\sin\alpha=(r/s)\sin\theta$ that $|r|\sim|s|$, so we must
also require that $r\gg1$. Thus, if $\tau\ll1$, the matrix elements of the $\cP
\cT$-symmetric Hamiltonian are large. Second, the result that $\tau=0$ does not
violate the uncertainty principle. Both Hermitian and non-Hermitian $\cP
\cT$-symmetric Hamiltonians share the properties that (i) the passage time is
given by $\pi\hbar/\omega$, and (ii) $\Delta H\leq\omega/2$.

To summarize, the key dif\/ference between the Hermitian and the non-Hermitian
case is that $\left(1\atop0\right)$ and $\left(0\atop1\right)$ are orthogonal in
the Hermitian case, but they have separation $\pi-2|\alpha|$ in the $\cP
\cT$-symmetric case. This is because the Hilbert space metric of a $\cP\cT$
quantum theory depends on $H$. By choosing the parameter $\alpha$ properly, we
create a wormhole-like ef\/fect in Hilbert space. That is, we f\/ind a path in
Hilbert space from the initial state vector $|\psi_I\rangle$ to the f\/inal state
vector $|\psi_F\rangle$ that is shorter than the Hermitian path. This is
analogous to f\/inding a wormhole in coordinate space. In short, what we have done
here is to construct a ``wormhole'' in Hilbert space.

\subsection*{Acknowledgements}

The author receives f\/inancial support from the U.S. Department of Energy.

\pdfbookmark[1]{References}{ref}
\LastPageEnding
\end{document}